\documentclass[doublecol]{epl2} 

\usepackage[english]{babel}
\usepackage{times}
\usepackage{graphicx}
\usepackage{amsfonts}
\usepackage{psfrag}
\usepackage{verbatim}
\usepackage{color}

\begin{document}

\title{Thermal quenches in the stochastic Gross-Pitaevskii equation:\\
morphology of the vortex network}
\shorttitle{Morphology of the vortex network}

\author{Michikazu Kobayashi\inst{1} and Leticia F. Cugliandolo\inst{2,3}} 

\institute{                    
  \inst{1} Department of Physics, Kyoto University, Oiwake-cho, Kitashirakawa, Sakyo-ku, Kyoto 606-8502, Japan \\
  \inst{2} Universit\'e Pierre et Marie Curie
  - Paris 6, Laboratoire de Physique Th\'eorique et Hautes Energies,
  4, Place Jussieu, Tour 13, 5\`eme \'etage, 75252 Paris Cedex 05,  France
 \inst{3} Kavli Institute of Theoretical Physics, University of California, Santa Barbara, Santa Barbara, CA 93106, USA
}


\abstract{
We study the evolution of 3d weakly interacting bosons at finite chemical potential with the stochastic Gross-Pitaevskii equation. We fully characterise 
the vortex network in an out of equilibrium.  At high temperature the filament statistics are the ones of fully-packed loop models. The vortex tangle undergoes a 
geometric percolation transition within the thermodynamically ordered phase. After infinitely fast quenches across the thermodynamic critical point deep into the 
ordered phase, we identify a first approach towards the critical percolation state, a later coarsening process that does not alter the fractal properties of the long 
vortex loops, and a final approach to equilibrium. Our results are also relevant to the statistics of linear defects in type II superconductors, magnetic materials and 
cosmological models.
}


\maketitle



Three dimensional complex field theories with continuous symmetry breaking are used to describe phase 
transitions in a host of physical systems including 
superfluid $^4$He~\cite{Ahlers,Griffin,Minnhagen},
type II superconductors~\cite{Minnhagen,Vinokur}, 
nematic liquid crystals~\cite{deGennesProst}, 
magnetic materials~\cite{Magnets}, weakly interacting bosons~\cite{Blaizot}, and 
the early universe~\cite{HindmarshKibble,Vilenkin}. 
Such phase transitions lead to the formation of topological defects.
In recent years, an impressive theoretical, numerical and experimental 
effort has been devoted to the measurement of the density of defects after
slow quenches through second order phase transitions~\cite{delCampo}. However, very little is known 
about the size distribution, geometric properties and spatial organisation of the 
defects inherited from fast and slow quenches. Such questions have been addressed in 
spin models, and mostly in two dimensions~\cite{Arenzon07,Sicilia07}. 

Of particular interest are linear  defects, 
be them vortices, disclinations, or cosmic strings supported by
$3d$ field theories with global U(1) symmetry that capture, 
in different limits, the systems mentioned in the 
first paragraph. In this Letter we study the dynamics engendered by 
the stochastic Gross-Pitaevskii equation for
weakly interacting bosons~\cite{Gardiner}
from the point of view of the vortex tangle. \textcolor{black}{
We show that the number density of vortex lengths is the one of fully packed models at high $T$
and we fully characterise its time-dependence after sudden quenches.}
(Relativistic or underdamped extensions
yield similar results apart from short-time differences~\cite{Kobayashi-long}.) 

The stochastic dynamics of the space-time dependent complex scalar field $\psi({\mathbf x},t)$ 
are ruled by the  Langevin equation 
\begin{equation}
(2 i \mu - \gamma_{\rm L}) \dot{\psi} 
 = - \nabla^2 \psi + g(|\psi|^2 - \rho) \psi 
 - \sqrt{\gamma_{\rm L} T} \xi
 .
\label{eq:under-damped-non-relativistic}
\end{equation}
The complex noise $\xi$ has Gaussian statistics with zero mean, $\langle \xi_a({\mathbf x},t) \rangle =0$, and 
$\langle \xi_a({\mathbf x},t) \xi_b({\mathbf x}', t') \rangle = \delta_{ab}
\delta({\mathbf x}-{\mathbf x}') \delta(t-t')$ for $a,b=1,2$ the real and imaginary components. 
$g$ and  $\rho$ are real parameters
in the Mexican-hat potential energy with degenerate minima at $|\psi|^2= \rho$, the zero 
temperature equilibrium density.
$\gamma_{\rm L}$ is a friction coefficient. In the following we measure the order parameter in units of $\sqrt{\rho}$, and 
space and time in units of the $T=0$ mean-field 
equilibrium correlation length, $x_0 \equiv 1 / \sqrt{g \rho}$, and correlation time, 
 $t_0 \equiv \gamma_{\rm L} / x_0^2$.
The chemical potential $\mu$, temperature $T$, and Langevin noise $\xi$ are measured in 
units of $t_0 / x_0^2$, $x_0^{-2}$, and $\sqrt{\rho / t_0}$, respectively  (we set the Boltzmann constant
to one hereafter). We place the fields in a  three spatial dimensional lattice and 
we solve the stochastic equation numerically with the lowest order Runge-Kutta method,
periodic boundary conditions (PBC), and parameters $g = 1$, 
$\mu = 0.5$ and $\gamma_{\rm L} = 1$. We \textcolor{black}{show data obtained using} discretization steps $\Delta x = 1$ and $\Delta t = 0.01$,
averaged over at least 100 noise realisations. \textcolor{black}{We checked that the equilibrium and dynamic 
results are not modified using smaller $\Delta x$ and $\Delta t$ when studied as functions of  $T/T_{\rm c}$
(the critical temperature $T_{\rm c}$ itself depends on $\Delta x$)}. 

Vortices are centred at points where the field vanishes and  its phase changes by $2\pi n$, with $n$ a non-vanishing integer, 
along a closed contour around  them. \textcolor{black}{A vortex element threads a square plaquette when  
the phase of the field on its four vertices changes by $\pm 2 \pi$. When more 
than one vortex element enter (and exit) a unit cell, we  use the 
stochastic scheme to recombine them. The length $l$ of a  vortex is the number of elements along it times $\Delta x$.}
 (We will discuss the effects of other reconnection 
rules in~\cite{Kobayashi-long}, see also~\cite{Kajantie,Bittner}.) For PBC
the lines are closed and can wrap around
the system. 

Equation~(\ref{eq:under-damped-non-relativistic})  takes the system to Gibbs-Boltzmann equilibrium and 
captures a second order {\it thermodynamic} phase transition
at $T_{\rm c}= 2.26 \pm 0.02$ for the parameters chosen. 
The critical exponents estimated numerically~\cite{Kobayashi-long} are very close 
to the ones expected from the $\epsilon$-expansion, and 
Monte Carlo simulations of the $3d$ XY model~\cite{Campostrini}.

\begin{figure}[h]
\centering
\includegraphics[width=0.7\linewidth]{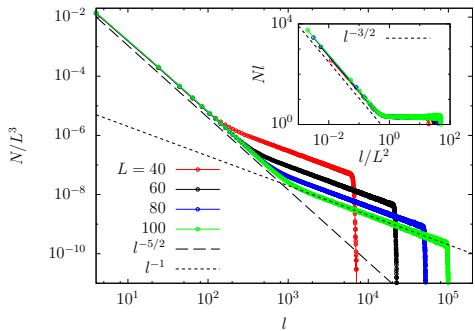}
\caption{(Colour online.)\label{fig:size-system-scale-inf}
Loop length number density at infinite temperature.
In the inset, the scaling plot.
}
\end{figure}

\textcolor{black}{A typical equilibrium state at very high temperature is such that two vortex elements enter and exist 
each unit cubic cell. }
The number density of vortex lengths normalised by the sample volume, $N(l)/L^3$, 
is shown in Fig.~\ref{fig:size-system-scale-inf} for various $L$. The data crossover 
from an algebraic decay with exponent $5/2$, as for Gaussian random 
walks~\cite{deGennes}, to a weaker decay with exponent $1$, the result for 
fully-packed oriented loop models with the two-in two-out constraint~\cite{Nahum}. 
Due to the Gaussian statistics short 
loops feel the size of the simulation box at scales ${\mathcal O}(L^2)$, as 
confirmed by the collapse of $l N$ {\it vs.} $l/L^2$ 
In short, 
\begin{eqnarray}
N(l)/L^3 \simeq 
\left\{
\begin{array}{ll}
l^{-5/2} 
&\qquad l \ll L^2 
\\
l^{-1} \ L^{-3}
&\qquad l \gg L^2
\end{array}
\right.
\label{eq:limits}
\end{eqnarray}
at $T \gg T_{\rm c}$.
Gaussian statistics were found in a $Z_3$ model for 
the initial condition of cosmological models~\cite{Vachaspati} and in random optical fields~\cite{Dennis08} observed in 
laser speckels and modelled with random wave superpositions. The cross-over to length-scales longer than 
$L^2$ was not discussed in these works.

\begin{figure}[h]
\vspace{0.2cm}
\centering
\includegraphics[width=0.7\linewidth]{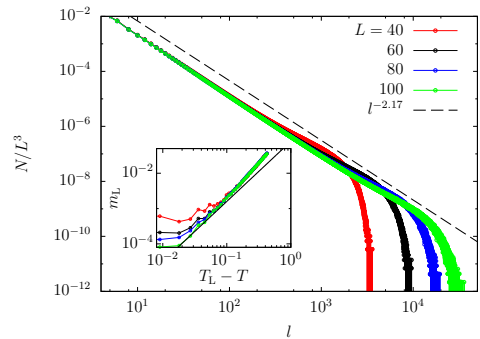}
\caption{(Colour online.)\label{fig:size-system-scale-TL}
Equilibrium line length number density at the line tension point
$T_{\rm L} = 0.98 \ T_{\rm c}$. Inset: $m_{\rm L}$ approaching $T_{\rm L}$ from below 
and the algebraic dependence with 
$\beta_{\rm L}=1.7$  (black line).
}
\end{figure}

The detailed geometric analysis of line ensembles  
has been very successful in, {\it e.g.}, polymer science~\cite{deGennes}. 
At a critical point, be it thermodynamic or geometric, the critical objects 
are fractal. Their Hausdorff dimension $D$ 
relates  the linear length along the loop, $l$ ($\gg l_0$ a microscopic length-scale), and the radius of the 
smallest sphere that contains the loop, $R$, 
as $l \simeq R^{D}$. 
In the thermodynamic limit,
\begin{equation}
N(l)/L^3 \simeq  l^{-\alpha_{\rm L}} \ e^{- l m_{\rm L}},
\end{equation}
with $m_{\rm L}$ the line tension and $\alpha_{\rm L}$ the  
`Fisher exponent'. (This form should be corrected
to capture finite size corrections.)
At criticality
$
m_{\rm L} \simeq |T-T_{\rm L}|^{\beta_{\rm L}}
$
with $\beta_{\rm L}$ another characteristic exponent.
$D$ and $\alpha_{\rm L}$ are linked by  
$
\alpha_{\rm L} = 1+d/D
$.

In equilibrium, the vortices undergo a 
 {\it geometric} transition between a `localised' phase 
with  finite loops only, and an `extended' phase in  which the length of 
a finite fraction of lines diverges in the thermodynamic limit. 
The  geometric threshold falls {\it within} 
the thermodynamically ordered phase, at $T_{\rm L} \simeq 0.98 \ T_{\rm c}$, 
where $m_L$ vanishes.  
In Fig.~\ref{fig:size-system-scale-TL}
we display $N(l)/L^3$ at $T_{\rm L}$. \textcolor{black}{An algebraic data fit yields $\alpha_{\rm L} = 2.17(2)$ and the dashed line} uses $\alpha_{\rm L} = 2.17$, 
accordingly  $D\simeq 2.56$, a value to be associated with self-seeking 
random walks.  In the inset, the variation of 
$m_{\rm L}$  approaching $T_{\rm L}$ suggests 
$\beta_{\rm L} \simeq 1.7$. 

\begin{figure}[h]
\centering
\includegraphics[width=0.69\linewidth]{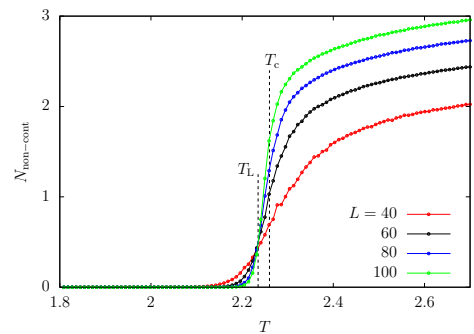}
\caption{(Colour online.)\label{fig:noncontract} 
Temperature dependence of the averaged number of non-contractible loops. The thermodynamic critical temperature, $T_{\rm c}$,
and the temperature for vanishing line tension, $T_{\rm L}$, are shown with vertical dotted lines. 
}
\end{figure}

We confirmed $T_{\rm L} <T_{\rm c}$ with the analysis of  other observables. 
We defined the size of a loop as the maximal
side of the parallelepiped covering the vortex in the three Cartesian 
directions, {\it i.e.}, the length that the string would have after smoothing out all small
scale irregularities. We then counted the loops with larger size than the system size.
Besides, we counted the non-contractible loops, {\it i.e.}, those with non-vanishing winding number in 
at least one of the three spatial directions. The two numbers  
behave as order parameters for the geometric transition.
Figure~\ref{fig:noncontract} displays the temperature dependence of the averaged $N_{\rm non-cont}$ 
together with $T_{\rm L}$ and 
$T_{\rm c}$ shown with vertical dotted lines.

\begin{figure}[h]
\vspace{0.2cm}
\centering
\includegraphics[width=0.7\linewidth]{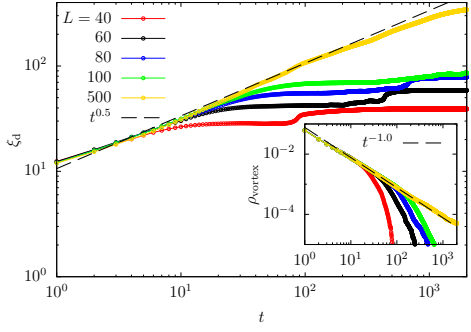}
\caption{(Colour online.)\label{fig:vortex-density} 
The dynamical correlation length $\xi_{\rm d}$ in systems with different sizes given in the key.
In the inset, the decay of the vortex density $\rho_{\rm vortex}$ in the same cases. 
}
\end{figure}

We conclude that the geometric threshold differs from the ther\-mo\-dynamic critical point. The same
fact was found in the $3d$ XY model~\cite{Kajantie} and the O(2) field theory~\cite{Bittner}.
These findings refute claims of the coincidence of the two transitions
in the context of cosmology~\cite{AntunesBettencourt,Schakel},  
in the field of superfluidity and type-II superconductivity~\cite{Nguyen,Camarda}, and
in general~\cite{Kohring}. See~\cite{NahumPRE}  for the field theory of the geometric transition.

\begin{figure}[h]
\vspace{0.2cm}
\centering
\includegraphics[width=0.7\linewidth]{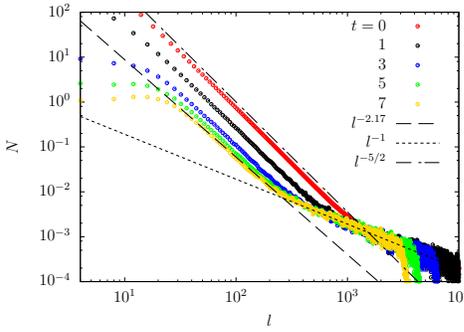}
\caption{(Colour online.)\label{fig:fast-size-initial-stochastic}
Early stages of evolution. 
Time dependent length number  $N(l,t)$ in a system with  linear  size  $L = 100$.
The algebraic decays with discontinuous black lines are explained in the 
 text. The fit in Eq.~(\ref{eq:guess}) is applied to the $t=7$ dataset
with $c_1 = 3.2$, $c_2 = 9.2 \times 10^{-3}$ and $n = 6$.}
\end{figure}

We now turn to the dynamics after instantaneous quenches. A system 
prepared in equilibrium at $T\gg T_{\rm c}$ is subsequently
evolved at $T=0$. For a sufficiently large $L$ the spatial averaged initial field 
is practically zero and the distribution of vortex lengths is very close to the 
one in Fig.~\ref{fig:size-system-scale-inf}.  After a transient,  the system enters a dynamic scaling regime
in which the space-time correlation, 
$C(r,t) \equiv \langle \psi^*({\mathbf x}, t) \psi({\mathbf x}', t) \rangle$ 
with $r \equiv |{\mathbf x} -{\mathbf x}'|$, or the dynamic structure 
factor,  scale as $C(r,t) \simeq$ $f(r/\xi_{\rm d}(t))$,
or $S(k,t) \simeq \xi^d_{\rm d}(t) \Phi(k \xi_{\rm d}(t))$, 
with a growing length $\xi_{\rm d}(t) = \gamma_{\rm v} t^{1/2}$, see Fig.~\ref{fig:vortex-density}.
This time-de\-pen\-dence was predicted in~\cite{BrayHumayun}
for continuous spin models with non-conserved order parameter. Consequently, 
 the vortex density decays as $t^{-1}$, see the inset. (We defined $\rho_{\rm vortex}$ as the number of 
 plaquettes with non-vanishing flux divided by their total number. Its `value' depends on $\Delta x$ but its time-dependence does not.) 
 Finite size effects,  
once analysed  {\it via}  $\xi_{\rm d}/L$ 
against $t/L^2$ yield  a first saturation to $\xi_{\rm d} \simeq 0.7 \, L$
at  $t\simeq 0.5 \, L^2$, before the final    approach to $\xi_{\rm d} = L$  at 
$t \simeq L^2$. 
Vortices inhibit long-range order in the entire space but, as soon as they disappear at the end of the first plateau, 
the system rapidly enters the state 
with large phase correlation.
This is the reason why $\xi_{\rm d}$ increases suddenly at $t\simeq 0.5 \, L^2$ and 
quickly approaches the system size $L$.

We focus now on the time-evolving 
number of rings with length $l$. We first describe the numerical 
data and we later give functional forms for the various regimes.

{\it Early regime}, Fig.~\ref{fig:fast-size-initial-stochastic}. 
At $t=1$,  $N$ is still very close to the initial one. Soon after,
a short-length regime  with positive slope starts to develop. To the right of the maximum, 
the algebraic decay is no longer  given by the Gaussian $5/2$ 
but by the  $\alpha_{\rm L}= 2.17$ of the critical geometric threshold. 
The time needed to reach this algebraic behaviour is $t_{\rm p} \simeq 7$ for $L=100$. The precise determination of the 
variation of $t_{\rm p}$ with $L$ is a hard task~\cite{Blanchard14,Tartaglia15} and goes beyond the scope of this Letter.
The weight of the longest length-scales 
remains the time-independent $l^{-1}$. The cross-over between the two power laws
occurs at a shorter $l$ than for the initial state
because the weight of the shorter loops has decreased by a factor of time.

\begin{figure}[h]
\includegraphics[width=0.68\linewidth]{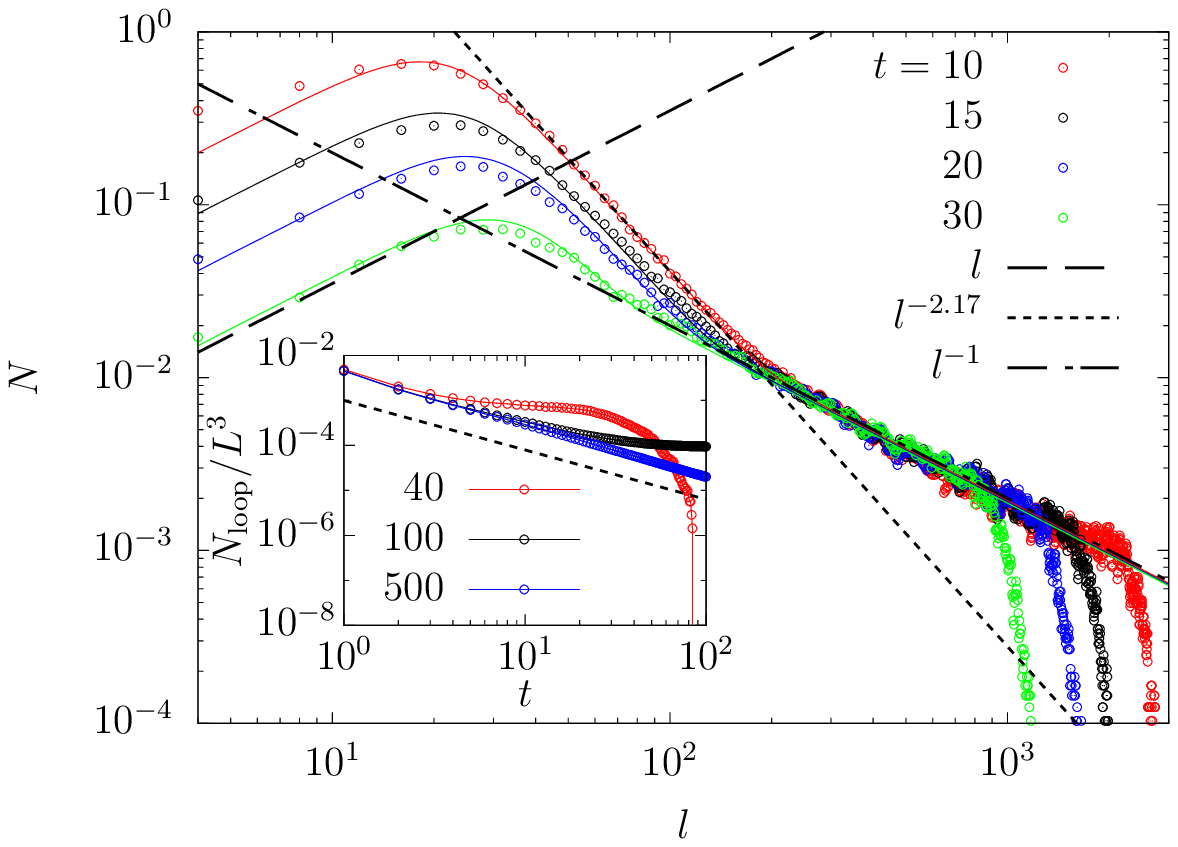}
\includegraphics[width=0.7\linewidth]{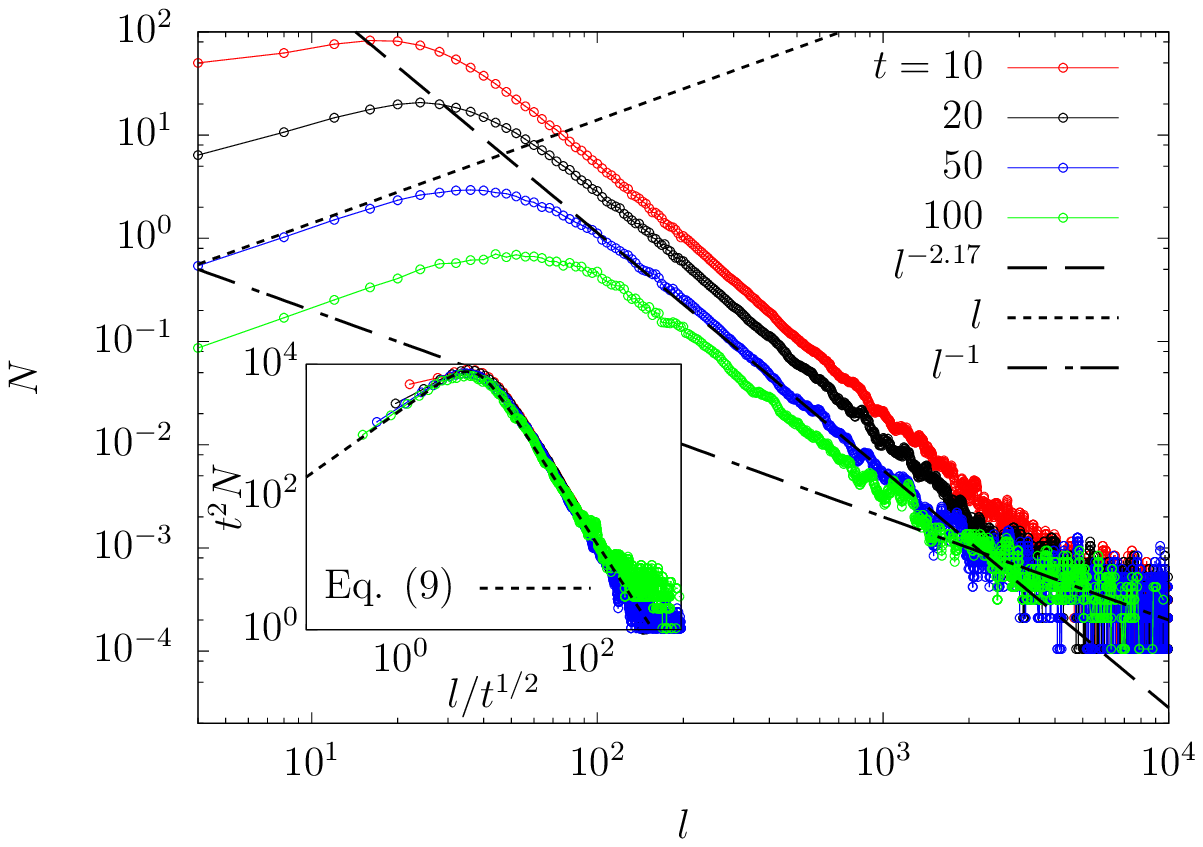}
\caption{(Colour online.)\label{fig:fast-size-power}
Dynamic scaling regime.
Time dependent length number $N(l,t)$  in systems with linear size $L = 100$ (above) and 
$L=500$ (below).
Circles and lines show the numerical results and their comparison to
Eq.~(\ref{eq:Pell-corr}) with $c_1 = 3.2$, $c_2 = 9.2 \times 10^{-3}$, $n = 6$, $\gamma_{\rm v} =0.85$, and $t_{\rm p} = 7$.
The discontinuous lines are $l^{-2.17}$,
 $l^{-1}$, and $l$. 
Upper panel inset: loop number density against time
together with $t^{-\zeta}$ with $\zeta=1.1$ (dotted line).
Lower panel inset: scaling plot $t^{2} N$  against $l/t^{1/2}$, and the 
functional form in Eq.~(\ref{eq:Pell-corr}).
}
\end{figure}

{\it Dynamic scaling regime}, Fig.~\ref{fig:fast-size-power}. 
Three scales  can be identified  for $L=100$ (above) and $L=500$ (below). 
At short lengths the curves 
are well represented by a linear dependence (dotted lines) until a maximum at a scale of the order of  
the growing correlation length $\xi_{\rm d}$. 
Beyond it a regime with weight $l^{-2.17}$ is seen in both panels \textcolor{black}{(data fits yield $\alpha_L = 2.17(9)$ and $\alpha_L=2.17(5)$)}. 
For the smaller system 
size, it is progressively erased and at  $t=30$ only the final $l^{-1}$ decay  remains. 
For the larger system size, the curves are parallelly 
translated down. The last $l^{-1}$ regime is also seen.
In the inset to the lower panel the scaling 
$t^{2}  N$ vs. $l/t^{1/2}$ describes the first two regimes 
quite accurately. 

\begin{figure}[h]
\includegraphics[width=0.68\linewidth]{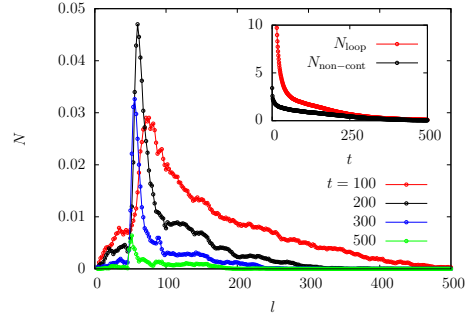}
\caption{(Colour online.)\label{fig:last}
Late epochs. Time dependent length number densities $N(l,t)$.
 Inset: averaged number of loops and
non-contractible loops vs. time, in  a system with $L=100$.
}
\end{figure}

{\it Late epochs}, Fig.~\ref{fig:last}.  At very long times just a few non-contractible long vortices 
remain, until they  break and  finally disappear, letting the system phase order.
The number of loops has a peak at 
$l\simeq L$. In the inset, the average number of loops, $N_{\rm loop}$,  and 
non-contractible loops, $N_{\rm non-cont}$, as functions of time. In equilibrium
these quantities vanish and, accordingly, the dynamic values approach zero.

A simple argument along the lines of the one used in~\cite{Arenzon07,Sicilia07}
to deal with the statistics of clusters in spin models leads to
an analytic - though approximate - expression for $N$.

The first remark is that the double algebraic decay of
$N$ at $t=0$  or $t_{\rm p}$  is 
well approximated by 
\begin{equation}
N(l, t) \simeq l^{-\alpha_{\rm L}} \ [c_1^{n}(t) + c_2^n(t) l^{(\alpha_{\rm L}-1) n}]^{1/n}
\; .
\label{eq:guess}
\end{equation}
A sharp  cross-over between the two power laws is obtained 
for large  $n$ ($n=6$ is sufficient, see Fig.~\ref{fig:fast-size-initial-stochastic}). 
The cross-over takes place at $l^*(t) \simeq [c_1(t)/c_2(t)]^{1/(\alpha_{\rm L}-1)}$.
If $\alpha_{\rm L}=5/2$, as for the initial state, $c_1(0) \propto L^3$ and $c_2(0)$ a finite constant ensure $l^*(0) \simeq L^2$ and the 
limit forms in Eq.~(\ref{eq:limits}). At $t_{\rm p}$ the weight of the loops with 
$l > l^*(t_{\rm p})$ is not modified with respect to the initial one, and $c_2(t_{\rm p}) = c_2$.
The total number of loops diminishes in time but remains $O(L^3)$ until the very late epochs, see the inset in 
the upper panel in Fig.~\ref{fig:fast-size-power}.
For $N$ as in (\ref{eq:guess}), to leading order in $L$, and 
ignoring constants, $N_{\rm loop}(t) = \int dl \, N(l,t) \simeq c_1(t) \int_{l_0}^{l^*} dl  \, l^{-\alpha_{\rm L}}
+ c_2(t) \int_{l^*}^{L^3} dl \, l^{-1} \simeq c_1(t) l_0^{1-\alpha_{\rm L}}$ 
that scales as $L^3$ if $c_1(t) =L^3 \overline c_1(t_{\rm p})$. 
According to Fig.~\ref{fig:fast-size-initial-stochastic},   $l^*(t_{\rm p}) < l^*(0)$. 

We assume that after the transient $t_{\rm p}$, 
the length of each vortex is reduced at the same rate as the \textcolor{black}{dynamic correlation length $\xi_d$} grows
\begin{equation}
l(t, l_{\rm p}) \simeq \gamma_{\rm v} \sqrt{t_{\rm a} - t}
\; , 
\end{equation}
with $\gamma_{\rm v}$ a parameter, 
$t_{\rm a} =  t_{\rm p} + l_{\rm p}^2 / \gamma_{\rm v}^2$ the annihilation time at which $l(t_{\rm a}) =0$, 
and $l_{\rm p}$ the length of the vortex ring at $t_{\rm p}$. We suppose that 
the vortices are sufficiently long and far apart that
they evolve independently of each other. Neglecting  the fact that they break up and disappear in the 
course of evolution we use
$N(l,t) \simeq \int dl_{\rm p} \, N(l_{\rm p}, t_{\rm p}) \, \delta (l - l(t, l_{\rm p}))$
to obtain
\begin{equation}
N(l,t) \simeq 
\frac{l \ N(\sqrt{\gamma_{\rm v}^2 (t-t_{\rm p}) + l^2}, t_{\rm p})}{\left[\gamma_{\rm v}^2 (t-t_{\rm p}) + l^2\right]^{1/2}}
 \; \label{eq:loop-distribution}. 
\end{equation}

Let us first focus on long length scales. From the numerator one estimates a 
crossover at a {\it dynamic length} $l^*(t)$, 
that is advected towards smaller scales as time evolves, 
\begin{eqnarray}
{l^*}(t) \approx \sqrt{(c_1(t_{\rm p})/c_2)^{1/(\alpha_{\rm L}-1)}- \gamma_{\rm v}^2 (t-t_{\rm p})}
\; ,
\end{eqnarray}
as observed in the numerical  data.
We recover
\begin{equation}
N(l,t) \simeq c_2 \ l^{-1} \qquad
l \gg l^*(t)
\end{equation}
independently of time. Instead, for  $l < l^*(t)$ we  need to 
correct (\ref{eq:loop-distribution}) to take into account  the annihilation of vortices with 
short length that implies $N_{\rm loop}(t) \simeq t^{-\zeta}$, 
see the inset in the upper panel in Fig.~\ref{fig:fast-size-power}, and the consequent 
reduction of the averaged length size $\langle l \rangle \simeq t^{-\zeta+1/2}$. We enforce this scaling heuristically, 
by simply multiplying Eq.~(\ref{eq:loop-distribution}) by $(\gamma_{\rm v}^2 t)^{-\zeta}$.
Proceeding in this way, and taking $t\gg t_{\rm p}$, 
\begin{equation}
(\gamma_{\rm v}^2 t)^{\zeta+\alpha_{\rm L}/2} \ N(l,t)  
\simeq 
\frac{ c_1(t_{\rm p}) \ \ l/ (\gamma_{\rm v}^2 t)^{1/2} }{\left[ 1+ l^2/(\gamma_{\rm v}^2 t) \right]^{(1+\alpha_{\rm L})/2}}
\; . 
\label{eq:Pell-corr}
\end{equation}
Finally, this regime can also be split in two 
\begin{eqnarray}
(\gamma_{\rm v}^2 t)^{\zeta+\frac{\alpha_{\rm L}}{2}} \  \frac{N(l,t)}{L^3}  
\simeq 
\left\{
\begin{array}{ll}
[l/(\gamma_{\rm v}^2 t)^{1/2}] 
& \;  l \ll \xi_{\rm d}(t) \;\;\;\;\;
\\
{[l/(\gamma_{\rm v}^2 t)^{1/2}]}^{-\alpha_{\rm L}} 
& \; l \gg \xi_{\rm d}(t) \;\;\;\;\;
\end{array}
\right.
\end{eqnarray}
For $\alpha_{\rm L} \simeq 2.17$ and $\zeta \simeq 1.1$, the exponent in the 
left-hand-side is close to $2$, the  value used in the inset  in Fig.~\ref{fig:fast-size-power} (lower panel).

\vspace{0.2cm}

In conclusion, we presented a full characterisation of the vortex network in and out of equilibrium. 
It should be possible to observe these static and dynamic vortex statistics in experimental
systems, \textcolor{black}{by, for example, examining the $3d$ vortex configurations in the turbulent state 
of a trapped Bose-Einstein condensate shortly after its expansion~\cite{Seman}.}

\acknowledgements
We thank I. Carusotto,  J. T. Chalker, M. Picco and N. Proukakis for useful discussions.
This research was  supported in part by NSF under 
Grant No. PHY11-25915, KAKENHI (Grant No. 26870295), 
Global COE Program ``the Physical Sciences Frontier'', the Photon Frontier Network Program, MEXT, Japan,
Grant-in-Aid for Scientific Research on Innovative Areas ``Fluctuation \& Structure'' (Grant No. 26103519)
from the Ministry of Education, Culture, Sports, Science, and Technology of Japan,
the JSPS Core-to-Core program ``Non-Equilibrium Dynamics of Soft-Matter and Information.'',
and the IRSES European Project ``SoftActive". LFC is a member of the Institut Universitaire de France.

\end{document}